\documentclass[conference,a4paper]{IEEEtran}
\IEEEoverridecommandlockouts
\usepackage{cite}
\usepackage{amsmath,amssymb,amsfonts}
\usepackage{algorithmic}
\usepackage{graphicx}
\usepackage{textcomp}
\usepackage{xcolor}
\def\BibTeX{{\rm B\kern-.05em{\sc i\kern-.025em b}\kern-.08em
    T\kern-.1667em\lower.7ex\hbox{E}\kern-.125emX}}
\begin{document}

\title{Learning to Learn to Compress}

\author{\IEEEauthorblockN{Nannan Zou\IEEEauthorrefmark{1}, Honglei Zhang\IEEEauthorrefmark{4}, Francesco Cricri\IEEEauthorrefmark{4}, Hamed R. Tavakoli\IEEEauthorrefmark{4},\\ Jani Lainema\IEEEauthorrefmark{4}, Miska Hannuksela\IEEEauthorrefmark{4}, Emre Aksu\IEEEauthorrefmark{4}, Esa Rahtu\IEEEauthorrefmark{1}}
\IEEEauthorblockA{\IEEEauthorrefmark{1}Tampere University, Tampere, Finland\\
\{nannan.zou, esa.rahtu\}@tuni.fi}
\IEEEauthorblockA{\IEEEauthorrefmark{4}Nokia Technologies, Tampere, Finland\\
\{honglei.1.zhang, francesco.cricri, hamed.rezazadegan\_tavakoli, jani.lainema, miska.hannuksela, emre.aksu\}@nokia.com}}

\maketitle

\begin{abstract}
In this paper we present an end-to-end meta-learned system for image compression. Traditional machine learning based approaches to image compression train one or more neural network for generalization performance. However, at inference time, the encoder or the latent tensor output by the encoder can be optimized for each test image. This optimization can be regarded as a form of adaptation or benevolent overfitting to the input content. In order to reduce the gap between training and inference conditions, we propose a new training paradigm for learned image compression, which is based on meta-learning. In a first phase, the neural networks are trained normally. In a second phase, the Model-Agnostic Meta-learning approach is adapted to the specific case of image compression, where the inner-loop performs latent tensor overfitting, and the outer loop updates both encoder and decoder neural networks based on the overfitting performance. Furthermore, after meta-learning, we propose to overfit and cluster the bias terms of the decoder on training image patches, so that at inference time the optimal content-specific bias terms can be selected at encoder-side. Finally, we propose a new probability model for lossless compression, which combines concepts from both multi-scale and super-resolution probability model approaches. We show the benefits of all our proposed ideas via carefully designed experiments. 
\end{abstract}


\section{Introduction}
\label{sec:intro}
In recent years, neural networks have been applied to image and video compression with promising results. The research community has followed mainly two directions. In one direction, one or more components of a traditional codec pipeline are implemented as neural networks, such as an in-loop filter \cite{Jia2019} or a post-processing filter \cite{ChaoDong2015}. In another direction, commonly referred to as \textit{end-to-end learned compression}, neural networks are used as the main components of the codec \cite{Chen2019a}. End-to-end learned methods have recently outperformed traditional codecs \cite{hu2020learning}. These systems typically follow the auto-encoder paradigm, where the encoder and decoder networks operate as non-linear transform and inverse transform, respectively, and a quantization step is used on the latent tensor output by the encoder \cite{Theis2017}. The quantized latents are then losslessly encoded, typically by using arithmetic coding with a learned probability model.  

A desired feature of machine learning tools is to generalize well to unseen content. However, current architectures and training methods still suffer from \textit{domain shift}, where the content type at test time is different from the content type considered during training. In the context of image compression, domain shift is likely to cause low rate-distortion (RD) performance at test time. Even when there is not significant domain shift between training and testing, the neural networks in the codec are not optimized on every unseen test image and thus content adaptation may still bring RD gains. One possible approach is to adapt the encoder neural network's parameters \cite{Aytekin2018a}. Another approach is to adapt the decoder network's parameters \cite{HongLam2019}. However, this latter approach would incur heavy bitrate overheads as the adapted parameters need to be signaled to the decoder. Yet another approach consists of adapting the latent tensor that is output by the encoder. As such adaptation processes is not meant for generalization, we refer to it also as \textit{benevolent overfitting}, or simply as overfitting.

However, neural networks in traditional end-to-end learned methods are trained for the only purpose of \textit{generalization} and not for maximizing the performance of the above overfitting approaches, thus generating a gap between training and inference stages. We argue that networks shall be trained specifically for \textit{both generalization and quick overfitting}, to reduce the number of overfitting iterations needed to achieve a certain gain.
The overfitting may be considered to be a form of few-shots learning, more specifically $1$-shot learning, where a model is adapted given a single example of an image.

In this paper, we describe L$^2$C, our end-to-end learned image compression system that we submitted to the \textit{2020 JPEG-AI Learning-based Image Coding Challenge}. Our main novel contributions are:
\begin{itemize}
\item A novel training paradigm for learned image compression, based on meta-learning for latent tensor overfitting. 
\item A novel probability model based on multi-scale progressive statistical model.
\item A novel and efficient strategy for generating an ensemble of content-specific decoders. 
\end{itemize}

\section{Prior Art}
\label{sec:priorart}
In end-to-end learned image codecs, the latent tensor output by the encoder is losslessly encoded by an entropy encoder given
the value distribution function estimated by the probability model.
The probability model is also referred to as an entropy bottleneck
layer as it constrains the amount of information that can be passed
through the latent tensor \cite{balle2018variational}. Element-wise
Gaussian distribution or other parametric distribution functions have
been proposed as the statistical model of the latent tensor \cite{minnen2018jointautoregressive,balle2018variational,balle2017endtoend}.
However, the assumption that the elements in the latent tenor are
independent of each other does not hold in practice. Experiments have
shown that the spatial correlation existing in the latent tensor degrades
the compression performance \cite{balle2018variational}\cite{lee2019contextadaptive}.
The authors of \cite{balle2018variational}\cite{lee2019contextadaptive}
address this problem by transferring extra information as a hyper-prior
to reduce this correlation. Instead of assuming the elements in the
latent tensor are independent and identically distributed, we propose
to use a powerful statistical model that is capable of capturing this
spatial correlation. The proposed multi-scale progressive
statistical model is based on previous multi-scale models used for lossless image compression
\cite{mentzer2019practical,cao2020lossless} and lossy image compression 
\cite{Zou2020}.

Overfitting the codec's parameters at inference time for compression has already been explored in the past. In \cite{Aytekin2018a}, the authors propose to overfit the encoder neural network to the test image. A similar technique was later used also in \cite{Aytekin2019}, \cite{Lu2020} for image and video compression, respectively. In \cite{HongLam2019}, the authors proposed to signal compressed weight-updates to the decoder. In order to encourage weight-updates to be robust to compression, a weight-update compression loss is used during the inference time overfitting. 
In \cite{Wang2020}, the authors propose an ensemble learning-based rate-distortion optimization (RDO) method to enhance end-to-end learned image compression. Multiple networks with same structure but different parametrization are used for the probability model in the lossless coding module. At inference time, the encoder selects the optimal probability model for each image block, and an index is signaled to the decoder. 
Latent tensor overfitting was proposed in \cite{Campos2019}, where the authors claim to achieve $0.5$ dB gain. 

As already mentioned in Section \ref{sec:intro}, a specialized training strategy may bring benefits in terms of overfitting performance. We see overfitting on test images as a $1$-shot learning problem. Many of the current state-of-the-art few-shots learning algorithms are based on Model-Agnostic Meta-Learning (MAML) \cite{Finn2017}, where training is split into an inner loop and an outer loop. In the inner loop, for each task in a batch of tasks, a model is adapted by gradient descent using training data (\textit{support set} in meta-learning terminology). In the outer loop, the adapted models are evaluated on validation data (\textit{query set}). The model is then updated by gradient descent on the evaluation losses. As the evaluation is performed using adapted models, the outer loop update computes second-order derivatives. In \cite{Nichol2018a}, first-order approximations of MAML were proposed, where the inner loop gradients are considered to be constant. Our proposed meta-learning strategy is inspired by MAML methods but it also presents fundamental differences. In particular, a "task" in MAML is for example a classification task, whereas in L$^2$C it is an image. For each "task", support set and query set include different images in MAML, whereas they include the same image in L$^2$C. This is because our adaptation is in fact an overfitting operation, whereas in MAML the adaptation aims at achieving generalization on the given task. In a variant of MAML, called CAML \cite{Zintgraf2019}, instead of adapting the parameters of a neural network, an additional context vector is adapted. Similarly, in L$^2$C the adaptation is performed on a tensor. However, CAML presents several differences to L$^2$C, in addition to those related to vanilla MAML. First, CAML aims at achieving reduced meta-overfitting, easier parallelization, and better interpretability, whereas L$^2$C aims at achieving better inference time overfitting. Second, the nature of the tensor is different: L$^2$C considers a latent tensor, output by another neural network. Third, the final goal of L$^2$C is image compression. In \cite{Jung2020}, a neural network is meta-learned so that it can be quickly pruned on the first video frame, and then used for tracking objects in subsequent frames. However, the target of compression is a neural network instead of the input content itself, and the final goal is object tracking instead of compression.  

\section{Methods}
\label{sec:methods}
\begin{figure*}[t]
\begin{centering}
\includegraphics[width=18cm]{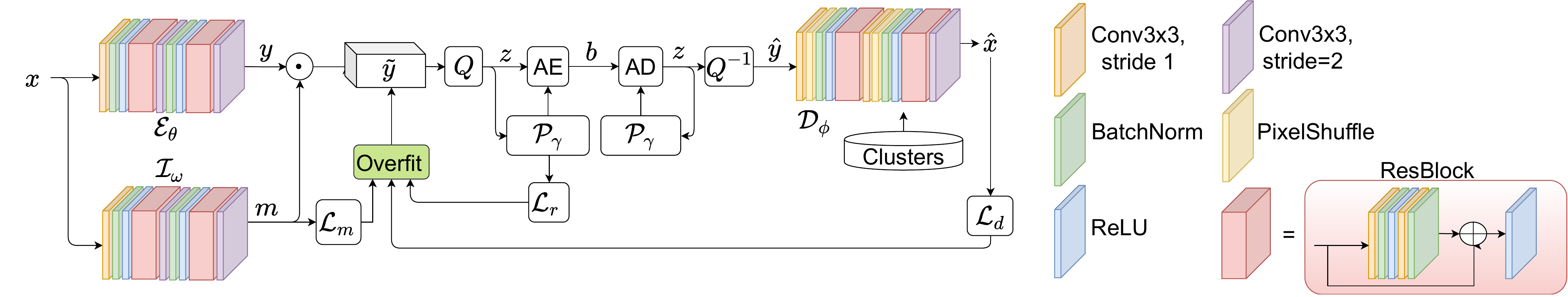}
\par\end{centering}
\caption{Overview of L$^2$C. $\mathcal{Q}$, $\mathcal{Q}^{-1}$, $\mathcal{P}_\gamma$, AE, AD, stand for quantization, dequantization, probability model, arithmetic encoder and arithmetic decoder, respectively.\label{fig:overview}}
\end{figure*}

In this section we describe in detail both the training phase and the inference phase of our end-to-end learned image compression codec L$^2$C. An overview is provided in Fig. \ref{fig:overview}.

The main components of L$^2$C are a fully-convolutional auto-encoder, an importance-map module for spatially-varying channel allocation, quantization and dequantization modules, a lossless coding module based on arithmetic coding and a learned probability model, and an ensemble of decoder-side clusters of content-specific bias terms. 

At encoder side, the input frame $x$ is projected into latent space by an encoder neural network $\mathcal{E}_{\theta}$ parametrized by weights $\theta$, obtaining the latent tensor $y \in \mathbb{R}^{\frac{H}{s},\frac{W}{s},c}$, where $H,W$ are the height and width of $x$, $s$ is the down-sampling factor of $\mathcal{E}_{\theta}$, $c$ is the number of filters in the last layer of $\mathcal{E}_{\theta}$. The output latent tensor is multiplied by a binary importance mask $m$ (see Section \ref{subsec:importance}) that zeros-out a spatially-varying number of channels: 
\begin{equation}
\tilde{y} = y \odot m = \mathcal{E}_{\theta}(x) \odot m,
\label{eq:masking}
\end{equation}
where $\odot$ indicates element-wise multiplication. The masked output is quantized into $z$ by uniform scalar quantization using $b$ bits and then entropy coded by an arithmetic encoder. In order to allow for back-propagating non-zero gradients, we use the straight-through estimator for quantization, as in \cite{Theis2017}. A learned multi-scale context model is used by the arithmetic codec to estimate the probability distribution of next symbols to encode/decode.
At decoder side, the entropy decoded bitstream is dequantized into $\hat{y}$ and input to a decoder neural network $\mathcal{D}_{\phi}$ parametrized by weights $\phi$, thus obtaining the reconstructed output image $\hat{x}$.

\subsection{Learned Spatially-varying Channel Masking}
\label{subsec:importance}
In order to allow the model to allocate a varying number of channels to different spatial areas of the encoded tensor $y$ (see \eqref{eq:masking}), we use an additional neural network $\mathcal{I}_\omega$ parametrized by weights $\omega$ and with similar architecture as $\mathcal{E}_{\theta}$. $\mathcal{I}_\omega$ takes as input the image $x$ and outputs an importance map $\tau \in \mathbb{R}^{\frac{H}{s}, \frac{W}{s},1}$ with elements in $[0,1]$. This map is then quantized with $\log_2 c $ bits and then expanded into a mask $m \in \mathbb{R}^{\frac{H}{s}, \frac{W}{s},c}$:
\begin{equation}
m_{i,j,k}=
\begin{cases}
1 & \text{if } k < c \tau_{i,j} \\
0 & \text{otherwise.}
\end{cases}
\label{eq:expansion}
\end{equation}
In order to encourage masked representations $\tilde{y}$ that have low entropy and thus be more easily predictable by our probability model for arithmetic coding, we use the following constraint in our training objective function:
\begin{equation}
\mathcal{M}(\tau) = \big\lvert \bar{\tau}-\zeta \big\lvert,
\label{eq:im_loss}
\end{equation}
where $\bar{\tau}$ is the mean value of $\tau$ and $\zeta$ is a constant representing the target average non-zero ratio in $m$ (and thus in $y_m$).

\subsection{Probability Model for Lossless Coding}
\label{subsec:probmodel}


\begin{figure*}[t]
\begin{centering}
\includegraphics[width=18cm]{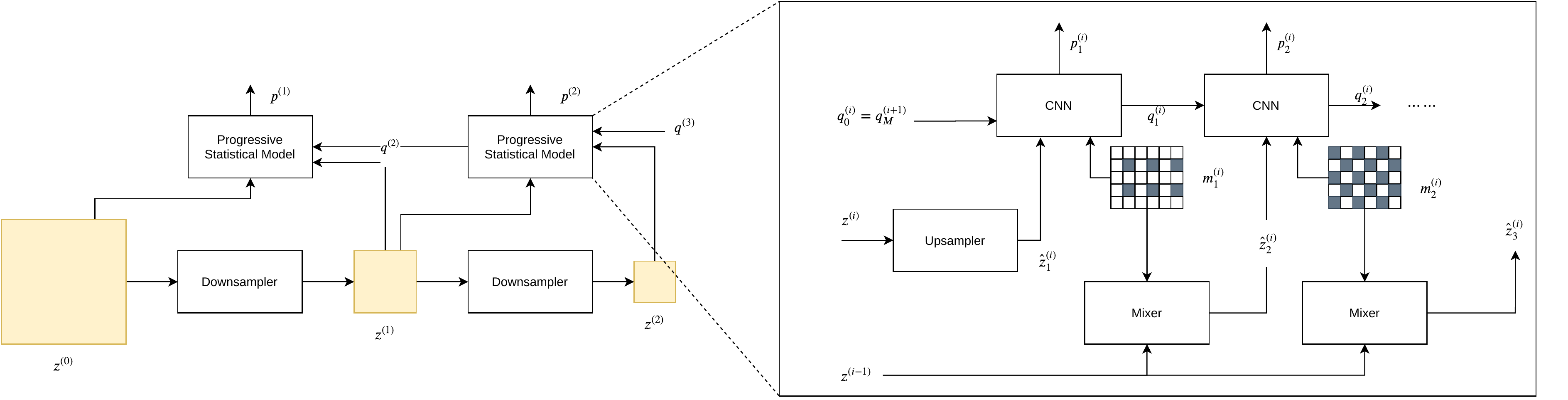}
\par\end{centering}
\caption{Architecture of the probability model that implements a multi-scale
progressive statistical model. \label{fig:Architecture-of_prob_model}}
\end{figure*}

Our proposed probability model implements a multi-scale progressive
statistical model based on the technologies used on lossless image
compression \cite{mentzer2019practical,cao2020lossless}. We first
downsample the latent tensor to a series of low-resolution representations
in multiple scales. The low-resolution representations are used as
a context in the statistical model for the elements in a high-resolution
representation. The system processes the multi-scale representations
backward. The representation in the last scale is sent without compression
since there is no context information for it. Other representations
are encoded/decoded based on the estimated distribution function using
the already processed low-resolution representations as input. This
procedure repeats until the input latent tensor is processed. 

Let $z^{(0)}$ be the latent representation to be modeled and $z^{(i)}$
be the low resolution representation of $z^{(0)}$ at scale $i$,
where $i=1,2,\cdots,M$ and $M$ is the number of scales. The joint
distribution function of elements in $z^{(0)}$ is defined by 

\begin{equation}
p(z^{(0)})=\left(\prod_{i=1}^{M-1}p\left(z^{(i-1)}\vert z^{(i)}\right)\right)p\left(z^{(M)}\right),\label{eq:multi-scale_distribution_model}
\end{equation}
where $p\left(z^{(M)}\right)$ is the distribution function of the
low-resolution representation at the last scale, which is assumed
to follow a uniform distribution. 

We choose the nearest neighbor downsampling method to avoid the extra
effort of encoding round-off errors as used in \cite{cao2020lossless}.
Note that this follows the design principle of the coupling layer
in Real-NVP algorithm \cite{dinh2017density}. With this design, the
elements at low-resolution representations are directly matched to
part of the elements in a high-resolution representation. 

To further improve the compression rate, we partition the elements
to be processed at each scale into multiple groups and process the
groups one by one. The groups that have been processed are added to
the context. Let $g_{j}^{(i)}$ be group $j$ at scale $i$, $B_{i}$
be the number groups at scale $i$, and $C_{j}^{(i)}=\left\{ g_{1}^{(i)},g_{2}^{(i)},\cdots,g_{j-1}^{(i)},z^{(i-1)}\right\} $
be the context for group $g_{j}^{(i)}$. The conditional distribution
$p\left(z^{(i-1)}\vert z^{(i)}\right)$ in \eqref{eq:multi-scale_distribution_model}
can be written as 
\begin{equation}
p\left(z^{(i-1)}\vert z^{(i)}\right)=\prod_{j=1}^{B_{i}}p\left(g_{j}^{(i)}\vert C_{j}^{(i)}\right).\label{eq:group_conditionay_prob}
\end{equation}

Let $z_{k}$ be a element to be processed in group $g_{j}^{(i)}$
and $N_{j}^{(i)}$ be the number of elements in group $g_{j}^{(i)}$.
We have 
\begin{equation}
p\left(g_{j}^{(i)}\vert C_{j}^{(i)}\right)=\prod_{k=1}^{N_{j}^{(i)}}p\left(z_{k}\vert C_{j}^{(i)}\right).\label{eq:pixel_conditional_prob}
\end{equation}
We assume $p\left(z_{k}\vert C_{j}^{(i)}\right)$ follows a mixture
of logistic distributions parameterized by the mixture weights, logistic
distribution means and scales \cite{salimans2016pixelcnn}. The parameters
are determined by a function modeled by a deep neural network with
$C_{j}^{(i)}$ as its input. 

The architecture of the proposed probability model is illustrated
in Figure \ref{fig:Architecture-of_prob_model}. The progressive statistical
model component takes $C_{j}^{(i)}$ as its input and outputs the
parameters of the value distribution functions and a context tensor
to be used at the next scale. The detailed structure of the progressive
statistical model component is shown on the right side of this figure.
For group $j$ at scale $i$, a deep neural network is used to calculate
$p_{j}^{(i)}$, which is the parameters for the value distribution
functions of the elements in $z_{j}^{(i)}$. The deep neural network
takes three inputs: $\hat{z}_{j}^{(i)}$ \textemdash{} a mixture of
the upsampled representation from the previous scale and the ground
truth tensor $z^{(i)}$; $m_{j}^{(i)}$ \textemdash{} a binary mask
that signals the ground truth availability of the elements in $\hat{z}_{j}^{(i)}$;
$q_{j-1}^{(i)}$ \textemdash{} a context tensor returned from the
previous step. This deep neural network also outputs $q_{j}^{(i)}$,
which is a context tensor to be used at the next step. 

At the training stage, the entropy of the input latent representation
$z^{(0)}$ is estimated using $p_{j}^{(i)}$ for all steps in all
scales. At the training stage, outputs $p^{(0)},p^{(1)},\cdots p^{(M-1)}$
are used to calculate the cross-entropy of the output tensors $z^{(0)},z^{(1)},\cdots,z^{(M-1)}$
and the sum of these cross-entropies is taken as the compression loss $\mathcal{L}_r(z)$.
At the encoding/decoding
stage, the value distribution functions for the elements to be encoded
are derived from $p_{j}^{(i)}$. Then the distribution functions are
given to the arithmetic encoder/decoder to encode/decode the elements
in $z_{j}^{(i)}$. This process starts from the last scale where the
elements are encoded/decoded using a uniform distribution function.
The encoding/decoding continues to every scale until all elements
in $z^{(0)}$ are processed.

\subsection{Meta-Learning}
\label{subsec:metalearning}
The training process is organized in two stages. In the first stage, a conventional training session is performed to achieve generalization, by using the following rate-distortion loss:
\begin{multline}
\label{eq:loss}
  \mathcal{L}(\hat{x}, x, z, \tau) = \lambda_{d_1} \mathcal{L}_{d_1}(\hat{x}, x) \\ + \lambda_{d_2} \mathcal{L}_{d_2}(\hat{x}, x) + \lambda_{d_3} \mathcal{L}_{d_3}(\hat{x}, x) + \lambda_{r} \mathcal{L}_{r}(z) + \lambda_m \mathcal{M}(\tau),
\end{multline}
where $\mathcal{L}_{d_1}$ is the negative multi-scale structural similarity (MS-SSIM) \cite{Wang2003a}, $\mathcal{L}_{d_2}$ is the mean-squared error (MSE), $\mathcal{L}_{d_3}$ is a perceptual loss, $\mathcal{L}_{r}$ is the rate loss provided by the probability model, $\mathcal{M}$ is the constraint on the importance map defined in \eqref{eq:im_loss}. $\lambda_{d_1}$, $\lambda_{d_2}$, $\lambda_{d_3}$, $\lambda_r$, $\lambda_m$ are scalar values that are determined empirically. Inspired by \cite{Johnson2016}, the perceptual loss is $\mathcal{L}_{d_3}=\big\lvert \mathcal{F}_{22}(x) - \mathcal{F}_{22}(\hat{x})\big\rvert_1 + \big\lvert \mathcal{F}_{43}(x) - \mathcal{F}_{43}(\hat{x})\big\rvert_1 $ where $\lvert \cdot\rvert_1$ is the $\ell_1$-norm, $\mathcal{F}_{22}$ and $\mathcal{F}_{43}$ are the ReLU 2-2 and ReLU 4-3 layers of VGG16 \cite{Simonyan2015} pretrained on ImageNet. The MSE loss ensures that pixel-wise differences are minimized, whereas MS-SSIM and the perceptual loss aim at improving the human-perceived quality.

The second stage is aimed at training the neural networks so that the performance of latent tensor overfitting is maximized at inference time. $\mathcal{E}_\theta$ and $\mathcal{I}_\omega$ need to be trained to output a latent tensor which can overfit quickly and effectively, and $\mathcal{P}_\gamma$ and $\mathcal{D}_\phi$ need to be trained to provide effective gradients during the overfitting process and to reconstruct high quality images from overfitted input latent tensors. We frame the problem as a few-shots learning problem, where the neural networks need to be trained to allow for overfitting the latent tensor in few iterations. As there's only one example image or \textit{shot}, this can be regarded as $1$-shot learning. The system is effectively  \textit{learning to overfit} for the task of image compression or, in other words, is \textit{learning to learn to compress} (L$^2$C). In practice, we fine-tune the neural networks trained in the first stage, by leveraging meta-learning and in particular MAML. The nature of the two nested loops used L$^2$C is different than in vanilla MAML.
In the inner loop, a latent tensor is overfitted for each image in a batch. In the outer loop, the neural networks are updated based on the average performance of the overfitted latent tensors over all images in the batch. 
\subsubsection{Inner-Loop}
\label{subsec:inner}
A batch of images $\textbf{x}=[x_1,x_2,..,x_B]$ is sampled from the training set. This is equivalent to a batch of tasks in vanilla MAML. For each image $x_i$, we perform an overfitting session. The initial step is to run one forward pass of the encoder-side networks, thus obtaining the initial latent tensor before quantization, $\tilde{y}_i^{(0)} =  \mathcal{E}_{\theta}(x_i) \odot m$.
The latent tensor is then updated by gradient descent for $n$ overfitting iterations and with learning rate $\alpha$. For each $k \in \{1,..,n\}$:
\begin{equation}
\tilde{y}_i^{(k+1)} = \tilde{y}_i^{k} - \alpha \nabla_{\tilde{y}_i} \mathcal{L}_{x_i}(\tilde{y}_i^{k}, \theta, \omega, \phi, \gamma)
\label{eq:latent_update}
\end{equation}
\subsubsection{Outer-Loop}
\label{subsec:outer}
The overfitted latent tensor $\tilde{y}_i^{(n)}$ of each image $x_i$ is used for computing an evaluation loss, $\mathcal{L}_{x_i}(\tilde{y}_i^{(n)}, \theta, \omega,\phi, \gamma)$. The neural networks' parameters are then updated by gradient descent with learning rate $\beta$:
\begin{multline}
\{\theta,\omega,\phi,\gamma\} = \{\theta,\omega,\phi,\gamma\} - \\ \beta \nabla_{\{\theta,\omega,\phi,\gamma\}} \sum\nolimits_{x_i\sim p(\mathcal{X})} \mathcal{L}_{x_i}(\tilde{y}_i^{(n)},\theta, \omega,\phi, \gamma).
\label{eq:outerupdate_enc}
\end{multline}
Assuming that $n=1$, this is equivalent to:
\begin{multline}
\{\theta,\omega,\phi,\gamma\} = \{\theta,\omega,\phi,\gamma\} - \\ \beta \nabla_{\{\theta,\omega,\phi,\gamma\}} \sum\nolimits_{x_i\sim p(\mathcal{X})} \mathcal{L}_{x_i}\Big(\tilde{y}_i^{(1)} - \\\alpha \nabla_{\tilde{y}_i} \mathcal{L}_{x_i}(\tilde{y}_i^{(0)}, \theta, \omega, \phi, \gamma), \theta, \omega,\phi, \gamma\Big).
\label{eq:outerupdate_enc2}
\end{multline}
\subsection{Adapting Decoders' Parameters}
\label{subsec:biases}
We propose to have multiple sets of overfitted decoder's parameters, from which the encoder can choose at inference time in order to adapt the decoding process to the test image. To this end, after training has completed, we overfit a subset of decoder's parameters to each 256x256 patch of the training set. In \cite{Lam2020}, the encoder adapts a neural network that is used as a post-processing filter within a conventional decoder. The authors found that updating only the bias terms is a good trade-off between gain in reconstruction quality and bitrate overhead incurred by signaling the updated weights. Thus, we overfit only the bias terms of the convolutional layers of $\mathcal{D}_\phi$. The obtained sets of overfitted bias terms are clustered by $k$-means into $255$ clusters. These clusters represent content-specific bias terms. At inference time, the optimal cluster can be signaled by using only $8$ bits. One index is reserved for signaling that no adaptation is needed and the default bias terms shall be used. 

\subsection{Rate Control and Inference Pipeline}
\label{subsec:inference}
The JPEG-AI Challenge defines the following 8 target bitrates: $\{r_0^T=2.0,r_1^T=1.5,r_2^T=1.0,r_3^T=0.75,r_4^T=0.5,r_5^T=0.25,r_6^T=0.12,r_7^T=0.06\}$, measured as bits-per-pixel (BPP). We achieve multiple bitrates by several strategies:
\begin{itemize}
\item We train four versions of our codec, $\{\mathcal{C}_0,\mathcal{C}_1,\mathcal{C}_2,\mathcal{C}_3\}$ with numbers of latent tensor's channels $\{c_0=8,c_1=6,c_2=3,c_3=1\}$ and numbers of quantization bits $\{b_0=8,b_1=8,b_2=4,b_3=4\}$. These models are trained to achieve target bitrates $2.0,0.75,0.12,0.06$, respectively, on the validation dataset.
\item At inference time, we select the most suitable number of quantization bits $b$ to achieve the desired target bitrate. 
\item Further adjustments to the bitrate are made via latent tensor overfitting, by choosing the weighting coefficients for the loss terms accordingly. 
\end{itemize}
At inference time, we perform the following operations. For each image $x_i$ and target bitrate $r_j^{T}$, with $j \in \{0,1,2,3,4,5,6,7\}$, the codec with closest target bitrate is selected. For example, for $r_0^T$, codec $\mathcal{C}_0$ is selected. Initially, the codec is run using $b=8$ quantization bits. If the achieved bitrate is higher than $r_0^T$, $b$ is decreased. Selection of the codec and of the number of bits is repeated until the bitrate is within the allowed margin for the target bitrate. Then, latent tensor overfitting is performed for further adjusting the bitrate and for optimizing the reconstruction quality. After overfitting, the optimal content-specific bias terms for $\mathcal{D}_\phi$ are selected for each image patch.

\section{Experiments}
\label{sec:experiments}
This section presents the experimental setup and results. The number of channels in the convolutional layers of $\mathcal{E}_\theta$, $\mathcal{I}_\omega$ and $\mathcal{D}_\phi$ is 192 except for the last layer, which is one 1 for $\mathcal{I}_\omega$ and 3 for $\mathcal{D}_\phi$. 
Training was performed on patches of size 256x256 extracted from the JPEG-AI dataset, using a batch-size of $12$. Adam optimizer was used for the first training stage and for the outer loop updates of the meta-learning fine-tuning stage. 
The initial training was performed for $200$ epochs, whereas the meta-learning fine-tuning was performed for $5$ epochs. In the inner-loop of the meta-learning fine-tuning, 4 overfitting iterations were used. The learning rates for the initial training, the outer loop updates and the inner loop overfitting were $0.0001,0.0001,0.1$, respectively. 

Table \ref{tab:rd} reports results for 4 images in the JPEG-AI test set.

\begin{table}[h!]
  \begin{center}
    \caption{Rate-distortion performance on 4 images from JPEG-AI test set: \textit{jpegai03} (ID \textit{03}), \textit{jpegai09} (ID \textit{09}), \textit{jpegai12} (ID \textit{12}), \textit{jpegai15} (ID \textit{15}). MSY is MS-SSIM computed on Y component. VMAF \cite{Z.Li} is computed on YUV444.}
    \label{tab:rd}
    \begin{tabular}{|c|c|c|c||c|c|c|c|} 
    \hline
      \textbf{ID} & \textbf{BPP} & \textbf{MSY} & \textbf{VMAF} & \textbf{ID} & \textbf{BPP} & \textbf{MSY} & \textbf{VMAF} \\
      \hline
      \hline
      $\textit{03}$ & $2.031$ & $0.9982$ & $94.32$ & $\textit{12}$ & $1.925$ & $0.9943$ & $93.04$ \\
      \hline
      $\textit{03}$ & $1.570$ & $0.9975$ & $93.86$ & $\textit{12}$ & $1.437$ & $0.9906$ & $90.99$ \\
      \hline
      $\textit{03}$ & $1.117$ & $0.9952$ & $91.02$ & $\textit{12}$ & $1.122$ & $0.9842$ & $88.39$ \\
      \hline
      $\textit{03}$ & $0.769$ & $0.9925$ & $86.42$ & $\textit{12}$ & $0.845$ & $0.9810$ & $87.29$ \\
      \hline
      $\textit{03}$ & $0.551$ & $0.9703$ & $81.12$ & $\textit{12}$ & $0.510$ & $0.9605$ & $83.11$ \\
      \hline
      $\textit{03}$ & $0.268$ & $0.9681$ & $69.96$ & $\textit{12}$ & $0.252$ & $0.9141$ & $79.22$ \\
      \hline
      $\textit{03}$ & $0.132$ & $0.9678$ & $56.79$ & $\textit{12}$ & $0.117$ & $0.8562$ & $61.04$ \\
      \hline
      $\textit{03}$ & $0.056$ & $0.9026$ & $38.85$ & $\textit{12}$ & $0.051$ & $0.7778$ & $30.28$ \\
      \hline
      \hline
      $\textit{09}$ & $2.247$ & $0.9938$ & $88.03$ & $\textit{15}$ & $1.969$ & $0.9980$ & $93.55$ \\
      \hline
      $\textit{09}$ & $1.581$ & $0.9860$ & $82.17$ & $\textit{15}$ & $1.489$ & $0.9973$ & $92.99$ \\
      \hline
      $\textit{09}$ & $1.148$ & $0.9827$ & $80.05$ & $\textit{15}$ & $1.065$ & $0.9941$ & $90.98$ \\
      \hline
      $\textit{09}$ & $0.644$ & $0.9717$ & $74.97$ & $\textit{15}$ & $0.673$ & $0.9920$ & $86.56$ \\
      \hline
      $\textit{09}$ & $0.473$ & $0.9573$ & $64.23$ & $\textit{15}$ & $0.535$ & $0.9563$ & $83.95$ \\
      \hline
      $\textit{09}$ & $0.217$ & $0.9231$ & $48.07$ & $\textit{15}$ & $0.249$ & $0.9610$ & $68.04$ \\
      \hline
      $\textit{09}$ & $0.128$ & $0.8822$ & $27.16$ & $\textit{15}$ & $0.103$ & $0.9553$ & $54.96$ \\
      \hline
      $\textit{09}$ & $0.068$ & $0.7326$ & $14.43$ & $\textit{15}$ & $0.066$ & $0.9297$ & $36.92$ \\
      \hline

    \end{tabular}
  \end{center}
\end{table}

The number of parameters in $\mathcal{E}_\theta$, $\mathcal{I}_\omega$, $\mathcal{P}_\gamma$, $\mathcal{D}_\phi$ are $1.6$M, $1.6$M, $1.9$M, $2.7$M, respectively.
We performed measurements for the inference time on a representative image, \textit{jpegai10}. For each patch, the encoding used $10$ overfitting iterations and then searched for the best biases among $255$ sets. Encoding and decoding took $290.2$ and $14.3$ seconds, respectively, on a workstation equipped with Intel\textsuperscript{\textregistered} Core\textsuperscript{\texttrademark} i7-7820X CPU @ 3.60GHz, 8 cores, 64GB RAM, Nvidia RTX 2080 Ti GPU with 12 GB memory, PyTorch 1.3.1, Python 3.7.  

We now compare our meta-learning fine-tuning to a baseline fine-tuning, in terms of latent tensor overfitting performance. For each number of iterations in $\{4, 10, 100\}$, each image and each fine-tuning method, we search for the best learning rate and report the results averaged over the test set in Table \ref{tab:ml}, in terms of percentage increase in average loss drop due to overfitting. As a concrete example, for image \textit{jpegai02} and 4 overfitting iterations, the baseline model increases MS-SSIM from $0.9721$ to $0.9727$ ($0.0006$ MS-SSIM gain) and BPP from $0.7666$ to $0.7694$ ($0.0028$ BPP increase), whereas L$^2$C model increases MS-SSIM from $0.9701$ to $0.9712$ ($0.0011$ MS-SSIM gain) and BPP from $0.6787$ to $0.6812$ ($0.0025$ BPP increase). 

\begin{table}[h!]
  \begin{center}
    \caption{Relative improvement in mean loss drop due to overfitting, for L$^2$C fine-tuning with respet to baseline fine-tuning. Num. iters is the number of overfitting iterations, LRs are the sets of tested learning rates.}
    \label{tab:ml}
    \begin{tabular}{|c|c|c|} 
    \hline
      \textbf{Num. iters} & \textbf{LRs} & \textbf{$\Delta$ mean loss drop} \\
      \hline
      \hline
      $4$ & $\{0.1,0.15,0.2\}$ & $4.58\%$ \\
      \hline
      $10$ & $\{0.01,0.02,0.05\}$ & $11.37\%$ \\
      \hline
      $100$ & $\{0.001,0.005,0.01\}$ & $9.96\%$ \\
      \hline
    \end{tabular}
  \end{center}
\end{table}
In Table \ref{tab:overfitting} we show the effectiveness of the latent tensor overfitting with our meta-learned model in decreasing the bitrate and in improving the reconstruction quality. For each of these two experiments, we randomly selected three images.

\begin{table}[h!]
  \begin{center}
    \caption{Latent tensor overfitting for six example images. BPP$_1$ and MS-SSIM$_1$ refer to pre-overfitting, whereas BPP$_2$ and MS-SSIM$_2$ refer to post-overfitting.}
    \label{tab:overfitting}
    \begin{tabular}{|c|c|c|c|c|} 
    \hline
      \textbf{Target BPP} & \textbf{BPP$_1$} & \textbf{BPP$_2$} & \textbf{MS-SSIM$_1$} & \textbf{MS-SSIM$_2$} \\
      \hline
      \hline
      $2.0\pm0.3$ & $2.363$ & $\mathbf{2.264}$ & $0.9966$ & $0.9952$ \\
      \hline
      $1.0\pm0.15$ & $1.262$ & $\mathbf{1.089}$ & $0.9868$ & $0.9876$ \\
      \hline
      $0.75\pm0.11$ & $0.950$ & $\mathbf{0.825}$ & $0.9917$ & $0.9908$ \\
      \hline
      \hline
      $0.25\pm0.0375$ & $0.157$ & $0.258$ & $0.9172$ & $\mathbf{0.9315}$ \\
      \hline
      $1.0\pm0.15$ & $0.950$ & $1.045$ & $0.9917$ & $\mathbf{0.9931}$ \\
      \hline
      $1.5\pm0.22$ & $1.384$ & $1.532$ & $0.9856$ & $\mathbf{0.9892}$ \\
      \hline
    \end{tabular}
  \end{center}
\end{table}
Finally, we evaluated our content-specific bias terms. The mean MS-SSIM gain is $0.0029$, averaged over all test images and bitrates.

\section{Conclusions}
\label{sec:conclusions}
We presented our end-to-end learned solution for the JPEG-AI challenge, which introduces a new training paradigm for learned image compression, based on meta-learning. The models are trained to allow for a more effective latent tensor overfitting at inference stage. Also, we proposed a novel probability model for the lossless coding module, and a mechanism for generating and using content-specific decoder-side parameters. In our experiments, we showed how these proposed idea bring benefits in terms of rate-distortion performance.

\bibliographystyle{IEEEtran}
\bibliography{bibliography_database_for_arxiv}

\end{document}